\documentclass{article}

\usepackage[english]{babel}

\usepackage[
    letterpaper,
    top=2cm,
    bottom=2cm,
    left=3cm,
    right=3cm,
    marginparwidth=1.75cm
]{geometry}

\usepackage{amsmath}
\usepackage{graphicx}
\usepackage{natbib}
\usepackage[colorlinks=true, allcolors=blue]{hyperref}

\usepackage{tikz}
\usetikzlibrary{positioning} 

\usepackage{authblk}

\title{\textbf{BreakGPT: Leveraging Large Language Models for Predicting Asset Price Surges}}

\author{Aleksandr Simonyan\thanks{\href{mailto:aleksandrsimonyan1996@gmail.com}{aleksandrsimonyan1996@gmail.com}}}

\date{} 

\begin{document}

\maketitle

\begin{abstract}
This paper introduces BreakGPT, a novel large language model (LLM) architecture adapted specifically for time series forecasting and the prediction of sharp upward movements in asset prices. By leveraging both the capabilities of LLMs and Transformer-based models, this study evaluates BreakGPT and other Transformer-based models for their ability to address the unique challenges posed by highly volatile financial markets. The primary contribution of this work lies in demonstrating the effectiveness of combining time series representation learning with LLM prediction frameworks. We showcase BreakGPT as a promising solution for financial forecasting with minimal training, and as a strong competitor for capturing both local and global temporal dependencies.
\end{abstract}

\section{Introduction}

The rapid advancements in deep learning have enabled the development of models capable of addressing a wide range of tasks across domains such as natural language processing, computer vision, and time series forecasting \citep{vaswani2017attention, devlin2018bert}. However, predicting financial market behavior, especially identifying price surges in cryptocurrency markets, remains a challenging problem due to the stochastic nature of financial data and the influence of external factors \citep{benth2003stochastic, cont2001empirical}. In recent years, Transformer-based models have demonstrated exceptional performance in time series forecasting by capturing long-range dependencies and temporal interactions \citep{vaswani2017attention, lim2021temporal, zhou2021informer}. Simultaneously, the emergence of large language models (LLMs) has paved the way for transfer learning applications in financial time series data, including cryptocurrency markets \citep{raffel2020exploring, liu2019roberta}.

This study introduces \textbf{BreakGPT}, an architecture that combines the strengths of LLMs and Transformer-based models for predicting cryptocurrency price surges. We evaluate multiple architectures, including a modified \textbf{TimeLLM} \citep{timellm2023} and \textbf{TimeGPT} \citep{timegpt2023}, assessing their effectiveness in detecting price surges in assets like Bitcoin and Solana \citep{nakamoto2008bitcoin, mcgovern2019financial}.

Key contributions of this study include: 
\begin{itemize} 
    \item Development of a modified \textbf{TimeLLM} architecture that adapts GPT-2 for time series prediction using domain-specific prompts and embeddings \citep{timellm2023, radford2019language}. 
    \item Implementation and comparison of various Transformer-based models that utilize attention mechanisms and convolutional layers to process financial time series data.
    \item Evaluation of these models on real-world cryptocurrency datasets, analyzing their effectiveness in predicting price surges.
\end{itemize}

We demonstrate that LLMs and Transformer-based architectures can significantly improve time series forecasting in cryptocurrency markets, outperforming traditional statistical models while addressing the challenges posed by volatile financial data.

\section{Related Work}

Transformer-based models have recently made significant progress in time series forecasting by capturing long-range dependencies and nonlinear patterns within data \citep{vaswani2017attention, lim2021temporal, zhou2021informer}. Large language models (LLMs) such as GPT-2 have further advanced this field by leveraging pre-trained models for temporal data \citep{radford2019language, brown2020language}. \textbf{TimeGPT} \citep{timegpt2023} is a notable example, applying LLMs to time series data and demonstrating strong performance across various domains like finance and healthcare by utilizing self-attention mechanisms to model temporal relationships effectively.

While TimeGPT has shown the potential of LLMs in time series forecasting, our work introduces domain-specific adaptations tailored to the unique challenges of cryptocurrency price prediction. Given the high volatility in cryptocurrency markets, specialized techniques are necessary. Our study compares the performance of LLM-based models, such as BreakGPT and \textbf{TimeLLM} \citep{timellm2023}, with more traditional Transformer-based architectures like ConvTransformer.

The original \textbf{TimeLLM} model, which forms the foundation for our approach, adapts LLMs for time series forecasting by incorporating temporal embeddings and modified input structures \citep{timellm2023}. In our model, we enhance the architecture introduced in TimeLLM for predicting asset price movements, making it more suitable for detecting significant price shifts in volatile financial markets. We also provide a thorough evaluation of BreakGPT alongside more traditional time-series architectures, such as ConvTransformer, to highlight their respective strengths in capturing short-term and long-term dependencies \citep{mcgovern2019financial}.

Several benchmark studies in financial time series prediction are relevant to our work. For instance, the study by \citet{LOB_DL_Study2022} investigates deep learning models using Limit Order Book (LOB) data for stock price prediction. However, due to the limited availability of LOB data in cryptocurrency markets, our work focuses on OHLC (Open-High-Low-Close) data, demonstrating that Transformer-based models can outperform traditional statistical methods in this context.

Additionally, DeepLOB \citep{LOB_DL_Study2022} leverages LSTM and CNN layers for modeling LOB data in high-frequency trading and serves as an important reference. While DeepLOB focuses on granular order book data, our study uses general-purpose Transformer-based models and TimeLLM, which are better suited for long-term predictions where such detailed LOB data is unavailable.

Building upon these works, our study pushes the boundaries of time series forecasting in cryptocurrency markets, which remain underexplored compared to traditional stock markets. We demonstrate the adaptability of both LLMs and Transformer-based architectures to handle highly volatile real-world data in these financial markets \citep{nakamoto2008bitcoin}.

\section{Data and Target Generation}

\subsection{Data Preparation}

The dataset used in this study consists of Solana cryptocurrency price data from February 1 to August 15. The time frame from July 15 to August 15 was reserved as the test set, allowing the model to be evaluated on recent market data. The original dataset only contained OHLC (Open-High-Low-Close) data, but we augmented it by engineering additional features such as the Simple Moving Average (SMA), Exponential Moving Average (EMA), Relative Strength Index (RSI), and Bollinger Bands (BB) \citep{murphy1999technical, edwards2012technical}. These additional features were included to capture more nuanced market behaviors, providing the model with richer input data for trend prediction. To further reduce noise and highlight significant trends, the data was resampled from 1-minute intervals to 5-minute intervals, ensuring a balance between detail and signal clarity. Volatility for each period was calculated using the standard deviation of log price differences within the interval \citep{bollerslev1986generalized}:

\[
\sigma = \sqrt{\frac{1}{n-1} \sum_{i=1}^{n} \left( \log\left(\frac{P_i}{P_{i-1}}\right) - \mu \right)^2 }
\]

where $\mu$ represents the mean log price change, and $n$ is the number of intervals.

\subsection{Target Creation}

The target creation process aims to identify key market patterns, specifically focusing on detecting Higher Highs (HH), Lower Lows (LL), Higher Lows (HL), and Lower Highs (LH) \citep{murphy1999technical}. By locating local maxima and minima over a rolling window of 5 periods, we ensure that only significant market fluctuations are captured. This method helps filter out minor price movements and emphasizes more substantial trends. The detection of these extrema is based on examining the relationships between consecutive price points, ensuring that meaningful market reversals and trends are recognized for further analysis.

\[
\text{HH:} \quad P_i > P_{i-1} > P_{i-2}
\]
\[
\text{LL:} \quad P_i < P_{i-1} < P_{i-2}
\]

where $P_i$ represents the price at time $i$. We employ $K=2$ to ensure that two consecutive peaks or troughs confirm the trend \citep{edwards2012technical}.

Additionally, we apply a volatility filter where the price at the end of a given period must be at least 0.5\% greater than the initial price to qualify as a significant price change \citep{bollerslev1986generalized}:

\[
\Delta P = \frac{P_{\text{end}} - P_{\text{start}}}{P_{\text{start}}}
\]
\[
\text{An uptrend is identified if } \Delta P > 0.005
\]

This condition ensures that only significant price movements are considered for trend detection. For the purpose of binary classification in this work, we simplify the target labels to classify Uptrend (1) vs No Uptrend (0).

\section{Model Architectures and Results}

In this section, we present three models designed for predicting uptrends in financial time series data: the \textit{Simple Transformer} (used as a baseline), the \textit{ConvTransformer}, and \textit{BreakGPT}. These models address unique challenges such as capturing local patterns, handling volatility, and learning long-term dependencies in time series data.

The \textit{Simple Transformer}, used as a baseline, includes the basic architecture of embedding layers, multi-head attention, positional encoding, and fully connected layers. Although this model can capture long-term dependencies, it struggles with local temporal patterns in volatile financial data and serves as a point of comparison for the more advanced architectures \citep{vaswani2017attention, wu2020transformer}.

\subsection{ConvTransformer}

The \textit{ConvTransformer} enhances the Simple Transformer by incorporating a 1D convolutional layer before the transformer encoder. This enables the model to better capture both short-term and long-term patterns in volatile data \citep{wu2020transformer}.

\textbf{Key Components}:
\begin{itemize}
    \item \textbf{Input Projection}: Maps the raw time series input into a higher-dimensional space.
    \item \textbf{1D Convolution}: Captures short-term temporal patterns in the time series data, acting as a local feature extractor \citep{wu2020transformer}.
    \item \textbf{Residual Connections and SILU Activation Functions}: Enhance the flow of gradients and model expressiveness \citep{he2016deep}.
    \item \textbf{Positional Encoding}: Preserves the temporal order of the input data, ensuring that the transformer encoder is aware of the sequence's structure \citep{vaswani2017attention}.
    \item \textbf{Multi-Head Attention and Transformer Encoder}: Captures long-term dependencies and interactions within the input sequence \citep{vaswani2017attention}.
\end{itemize}

\subsection{BreakGPT}

The \textit{BreakGPT} is a modified version of GPT-2 \citep{radford2019language} adapted for time series classification. This model benefits from a prompt-based approach, where a predefined prompt guides the model's attention toward detecting sharp upward movements in financial time series data.

\textbf{Key Components}:
\begin{itemize}
    \item \textbf{Input Projection Layer}: Projects the input time series features into a higher-dimensional space to match the GPT-2 embedding dimensions.
    \item \textbf{Prompt}: A custom prompt guides the model's focus on detecting upward trends. The prompt used is:
    \begin{quote}
        \textit{"Predict if the current sequence signals the start of a sharp upward movement at the end."}
    \end{quote}
    \item \textbf{GPT-2 Encoder}: Processes both the projected time series data and the prompt using GPT-2's self-attention mechanism to capture long-term dependencies in the time series \citep{radford2019language}.
\end{itemize}

\subsection{Performance Evaluation}

All three models were evaluated based on their ability to detect uptrends (class 1) using precision, recall, F1-score, and accuracy. Due to class imbalance, the F1-score for class 1 was prioritized. The table below summarizes the performance of each model:

\begin{table}[h!]
\centering
\caption{Classification Performance of Models}
\footnotesize
\begin{tabular}{|c|c|c|c|c|c|c|}
\hline
\textbf{Model} & \textbf{Class} & \textbf{Precision} & \textbf{Recall} & \textbf{F1-Score} & \textbf{Accuracy} & \textbf{Overall F1-Score} \\
\hline
Simple Transformer & No Uptrend & 0.99 & 0.96 & 0.97 & 0.95 & 0.55 \\
                   & Uptrend    & 0.08 & 0.24 & 0.12 &       &      \\
\hline
ConvTransformer    & No Uptrend & 0.99 & 0.92 & 0.95 & 0.91 & 0.58 \\
                   & Uptrend    & 0.12 & 0.65 & 0.20 &       &      \\
\hline
BreakGPT           & No Uptrend & 0.99 & 0.96 & 0.98 & 0.95 & 0.57 \\
                   & Uptrend    & 0.11 & 0.31 & 0.16 &       &      \\
\hline
\end{tabular}
\label{tab:classification_performance}
\end{table}

\subsection{Discussion}

The \textit{Simple Transformer}, while serving as a baseline, struggled to capture the necessary patterns for predicting price uptrends in volatile financial data, achieving a low F1-score for class 1 even after more than 100 epochs. In contrast, the \textit{ConvTransformer} showed a significant improvement by integrating 1D convolutional layers, residual connections, and SILU activation functions. These enhancements enabled the model to effectively capture both short-term fluctuations and long-term dependencies, leading to an F1-score of 0.20 for class 1—a notable advancement in predicting uptrends. The \textit{BreakGPT} model demonstrated strong potential in just 10 epochs, performing close to the ConvTransformer. The prompt-based approach proved valuable in guiding the model to key features, and we anticipate further gains by using more advanced LLMs to enhance its predictive performance.

\section{Conclusion and Future Work}

In this study, we evaluated three models for predicting price surges in cryptocurrency markets. The ConvTransformer performed well by capturing both short-term and long-term dependencies, particularly in volatile datasets. The BreakGPT model, despite minimal training, demonstrated considerable promise with its prompt-based learning approach, showing strong potential for further improvements with more advanced models.

Future work will explore more sophisticated LLM architectures to further enhance predictive accuracy. Additionally, addressing class imbalance through advanced techniques like oversampling, class weighting, or ensemble learning will be key to refining model performance, especially for detecting uptrends in imbalanced financial datasets.

\bibliographystyle{apalike}
\bibliography{sample}

\newpage
\section*{Appendix}

In this appendix, we present the architecture diagrams for the two advanced models developed in this work: the \textit{ConvTransformer} and \textit{BreakGPT}. Each diagram provides a visual overview of the key components and the flow of data through the model. Additionally, we offer further explanations for each model's architecture and functionality.

The \textit{ConvTransformer} combines 1D convolutional layers with a Transformer encoder to capture both local and global patterns in the financial time series. The 1D convolution helps in extracting short-term patterns, while the Transformer encoder captures long-term dependencies. This hybrid approach allows the model to effectively balance learning from both local and global temporal patterns, making it especially suitable for volatile financial markets where short-term fluctuations and long-term trends are equally important for accurate predictions.

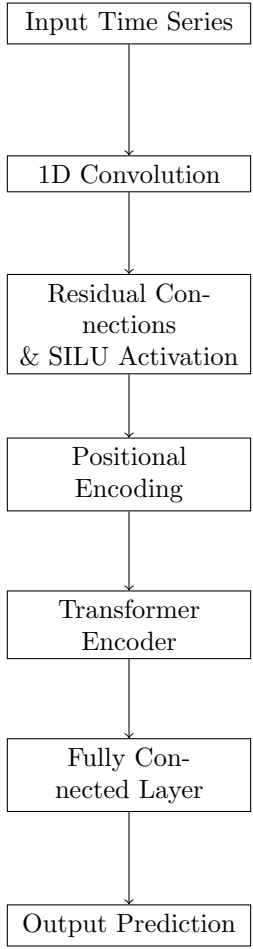
\begin{figure}[h!]
    \centering
    \begin{tikzpicture}[node distance=2cm]
        \node (input) [rectangle, draw, text centered, text width=3cm] {Input Time Series};

        \node (conv) [rectangle, draw, below of=input, text centered, text width=3cm] {1D Convolution};

        \node (residual) [rectangle, draw, below of=conv, text centered, text width=3cm] {Residual Connections \\ \& SILU Activation};

        \node (posenc) [rectangle, draw, below of=residual, text centered, text width=3cm] {Positional Encoding};

        \node (transformer) [rectangle, draw, below of=posenc, text centered, text width=3cm] {Transformer Encoder};

        \node (fc) [rectangle, draw, below of=transformer, text centered, text width=3cm] {Fully Connected Layer};

        \node (output) [rectangle, draw, below of=fc, text centered, text width=3cm] {Output Prediction};

        \draw[->] (input) -- (conv);
        \draw[->] (conv) -- (residual);
        \draw[->] (residual) -- (posenc);
        \draw[->] (posenc) -- (transformer);
        \draw[->] (transformer) -- (fc);
        \draw[->] (fc) -- (output);
    \end{tikzpicture}
    \caption{ConvTransformer Architecture}
    \label{fig:convtransformer}
\end{figure}

The architecture begins with the input time series being mapped to a higher-dimensional space through an embedding layer. The 1D convolutional layer processes the input, capturing local temporal dependencies such as sudden price fluctuations \citep{wu2020transformer}. Residual connections and SILU activation functions are used to enhance the model's expressiveness and gradient flow \citep{he2016deep}. Positional encodings are then added to ensure that the temporal order is preserved as the data moves through the Transformer encoder \citep{vaswani2017attention}. Finally, the Transformer encoder captures long-term dependencies before passing the processed data through a fully connected layer to produce a prediction.

The \textit{BreakGPT} model is a modified GPT-2 architecture tailored for time series classification. The model uses a prompt-based approach, where a predefined prompt helps guide the attention mechanism of GPT-2 to focus on detecting sharp upward movements in the time series data. The prompt and the input time series are concatenated and passed through the GPT-2 encoder, which captures long-term dependencies before making a final prediction \citep{radford2019language}.

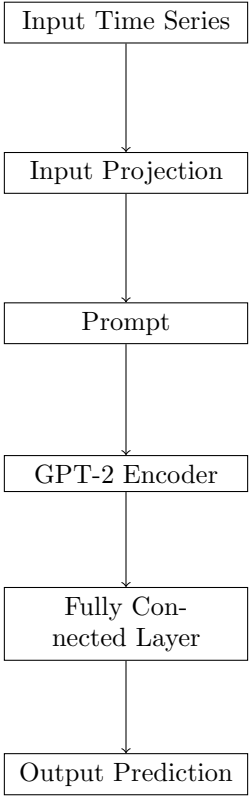
\begin{figure}[h!]
    \centering
    \begin{tikzpicture}[node distance=2cm]
        \node (input) [rectangle, draw, text centered, text width=3cm] {Input Time Series};

        \node (projection) [rectangle, draw, below of=input, text centered, text width=3cm] {Input Projection};

        \node (prompt) [rectangle, draw, below of=projection, text centered, text width=3cm] {Prompt};

        \node (gpt) [rectangle, draw, below of=prompt, text centered, text width=3cm] {GPT-2 Encoder};

        \node (fc) [rectangle, draw, below of=gpt, text centered, text width=3cm] {Fully Connected Layer};

        \node (output) [rectangle, draw, below of=fc, text centered, text width=3cm] {Output Prediction};

        \draw[->] (input) -- (projection);
        \draw[->] (projection) -- (prompt);
        \draw[->] (prompt) -- (gpt);
        \draw[->] (gpt) -- (fc);
        \draw[->] (fc) -- (output);
    \end{tikzpicture}
    \caption{BreakGPT Architecture}
    \label{fig:breakgpt}
\end{figure}

This architecture leverages GPT-2's self-attention mechanism to capture long-term dependencies in sequential data \citep{radford2019language}. The use of a prompt fine-tunes the model’s focus, helping it detect significant upward trends. The combined embeddings of the input time series and the prompt are passed through the GPT-2 encoder. The final hidden states are then passed through a fully connected layer to produce the output prediction.

Both models demonstrated their suitability for time series classification, particularly for predicting volatile financial market trends. The ConvTransformer excels at capturing local and global patterns through its hybrid approach, while BreakGPT leverages GPT-2’s attention mechanisms to detect sharp price movements.

\end{document}